\documentclass[utf8]{FrontiersinVancouver} 

\usepackage{url,hyperref,lineno,microtype,subcaption}
\usepackage[onehalfspacing]{setspace}

\def\keyFont{\fontsize{8}{11}\helveticabold }
\def\firstAuthorLast{Bendazzoli {et~al.}} 
\def\Authors{Simone Bendazzoli\,$^{1,2,*}$, Antonios Tzortzakakis\,$^{2,5,6}$,Andréas Abrahamsson\,$^{5}$, Björn Engelbrekt Wahlin\,$^{7,8}$, Örjan Smedby\,$^{1}$, Maria Holstensson\,$^{2,3,5}$,  and Rodrigo Moreno\,$^{1,4}$}
\def\Address{$^{1}$KTH, Royal Institute of Technology, Department of Biomedical Engineering and Health Systems, Sweden\\
$^{2}$Karolinska Institutet, Department of Clinical Sciences, Intervention and Engineering, Sweden
\\
$^{3}$Division of Medical Radiation Physics, Department of Physics, Stockholm University, Stockholm, Sweden 
\\
$^{4}$Department of Neurobiology, Care Sciences and Society, Karolinska Institutet, Stockholm, Sweden 
\\
$^{5}$Department of Nuclear Medicine and Medical Physics, Section Nuclear Medicine Huddinge, Karolinska University Hospital, Stockholm, Sweden 
\\
$^{6}$Department of Nuclear Medicine and Medical Physics, Theranostics Trial Center, Karolinska University Hospital, Stockholm, SwedenStockholm, Sweden 
\\
$^{7}$Division of Hematology, Dept. of Medicine, Huddinge, Karolinska Institutet, Stockholm, Sweden
$^{8}$Medical Unit Hematology, Solna, Cancer, Karolinska University Hospital, Stockholm, Sweden
}
\def\corrAuthor{Hälsovägen 11C, Huddinge, Sweden}

\def\corrEmail{simben@kth.se}

\begin{document}
\onecolumn
\firstpage{1}

\title[Lymphoma Lesion Detection]{Anatomy-Aware Lymphoma Lesion Detection in Whole-Body PET/CT} 

\author[\firstAuthorLast ]{\Authors} 
\address{} 
\correspondance{} 

\extraAuth{}

\maketitle

\begin{abstract}

\section{}
Early cancer detection is crucial for improving patient outcomes, and 18F FDG PET/CT imaging plays a vital role by combining metabolic and anatomical information. Accurate lesion detection remains challenging due to the need to identify multiple lesions of varying sizes. In this study, we investigate the effect of adding anatomy prior information to deep learning-based lesion detection models.
In particular, we add organ segmentation masks from the TotalSegmentator tool as auxiliary inputs to provide anatomical context to nnDetection, which is the state-of-the-art for lesion detection, and Swin Transformer.  The latter is trained in two stages that combine self-supervised pre-training and supervised fine-tuning. The method is tested in the AutoPET and Karolinska lymphoma datasets. The results indicate that the inclusion of anatomical priors substantially improves the detection performance within the nnDetection framework, while it has almost no impact on the performance of the vision transformer.
Moreover, we observe that  Swin Transformer does not offer clear advantages over conventional convolutional neural network (CNN) encoders used in nnDetection.  These findings highlight the critical role of the anatomical context in cancer lesion detection, especially in CNN-based models. 


\tiny
 \keyFont{ \section{Keywords:}Medical Object Detection, PET/CT, Lymphoma, Anatomical priors, Swin Transformer, Retina U-Net, nnDetection} 
\end{abstract}

\section{Introduction}
Early detection of cancer is crucial for effective treatment and improves survival rates and patient quality of life by minimizing invasive procedures. 
Combined positron emission tomography (PET) and computed tomography (CT) (PET/CT) scans provide more comprehensive information about tumors, including anatomy and metabolism. Whole-body PET/CT detected new and unexpected primary malignant tumors in at least 1.2\% of cancer patients \citep{Ishimori2005-oh}, and PET/CT has demonstrated usefulness in early cancer diagnosis, potentially improving survival rates. 

In order to accurately identify and count multiple lesions in PET-CT, includinf smaller ones that might be overlooked by segmentation methods, automated methods for object detection have been proposed\citep{FALLAHPOOR2024107880,Albuquerque2025}.
By prioritizing the identification of all potential lesions, regardless of size or location, object detection can contribute to more comprehensive and accurate cancer evaluations on PET/CT imaging, contributing to earlier and more accurate cancer detection, ultimately leading to better patient outcomes.

Object detection in PET/CT remains particularly challenging in multifocal disease, where the goal is to accurately identify and count all lesions, including the small ones that segmentation methods may miss. This capability is crucial in disseminated cancers, where comprehensive lesion detection directly impacts clinical assessment \citep{Albuquerque2025,Gunasekaran2023}.

Detection methods have demonstrated versatility across cancer types, proving effective for both localized tumors and disseminated diseases such as lymphoma and metastatic melanoma. They are especially well suited for identifying organs and large lesions, as shown by frameworks like \texttt{nnDetection}~\citep{baumgartner2021nndetection} and Retina U-Net~\citep{retina-unet}.
In particular, \texttt{nnDetection} is considered the state-of-the-art for detection. This is a self-configuring framework for 2D and 3D medical object detection, designed to automatically adapt to new tasks and datasets. It has demonstrated competitive performance on several benchmarks, including the ADAM challenge for aneurysm detection and segmentation \citep{https://doi.org/10.5281/zenodo.3715848, Timmins2021}, and the LUNA16 challenge for pulmonary nodule detection \citep{1612.08012, Setio2017}. The framework has also been applied effectively to detect intracranial aneurysms in TOF-MRA and structural MRI \citep{2305.13398}.
\texttt{nnDetection} is built on the Retina U-Net architecture, which combines RetinaNet \citep{1708.02002}, a single-stage object detector network, with U-Net \citep{1505.04597} to incorporate pixel-wise segmentation supervision, thus enhancing detection accuracy. This architecture has demonstrated strong performance in various clinical applications, including lung cancer staging using PET/CT \citep{lungpet2022retinau} and kidney tumor detection in the KiTS21 dataset \citep{kits21retinau}, consistently outperforming conventional object detection models in terms of average precision.

In recent years, the incorporation of anatomical context into medical images has gained attention as a means of integrating the information available to deep learning models. This additional context allows networks to infer higher-level representations during training, leading to a more comprehensive understanding of the task. Several studies have demonstrated the value of anatomical priors in deep learning applications. For example, \cite{Bermudez2019} showed that integrating volumetric features of anatomical regions into a deep neural network for brain age prediction significantly enhanced performance compared to models trained solely on imaging data. Similarly, \cite{Jin2022} reported improved accuracy in medical object detection and classification when anatomical guidance was explicitly provided during training.
In the domain of segmentation, the In-Context Cascade Segmentation (ICS) framework \citep{2412.13299} achieved superior results in anatomically complex regions by enforcing anatomical consistency across image slices, outperforming baseline models.

More recently, the introduction of vision transformer (ViT)-based architectures into medical image analysis has led to notable advances in segmentation tasks. Among these, Swin UNeTR~\citep{2201.01266} replaces the traditional CNN encoder of U-Net~\citep{1505.04597} with a ViT~\citep{2010.11929}, allowing the model to capture long-range spatial dependencies. 
The mechanisms from ViTs are similar to the way radiologists assess PET/CT scans, where they 
consider spatial patterns and co-occurrence of lesions across regions, rather than evaluating them in isolation. For example, the presence of a lesion in one area may suggest a possible involvement in another area through lymphatic spread, while the absence of lesions in certain regions may be a negative predictive factor for others. Thus, using this interregional context could significantly improve automated detection in such a complex problem. Thus, it is relevant to assess whether ViTs are appropriate for lesion detection tasks.

In this study, we explore training strategies to improve deep learning performance for lymphoma lesion detection on whole-body PET/CT scans, with a particular focus on integrating prior anatomical knowledge. Given the disseminated nature of lymphoma and the spatial complexity of whole-body imaging, our approach builds upon state-of-the-art methods while introducing architectural adaptations to better capture long-range relationships between lesions in anatomically distant regions. To incorporate anatomical context into the detection process, we developed an anatomy-aware framework that integrates segmentation masks of 104 organs, obtained from TotalSegmentator \cite{Wasserthal2023}, directly into the training pipeline. This might help the model to better localize the lesions using structural information about the surrounding organs and tissues.
Moreover, to explore whether ViT-based encoders can improve lesion detection, we replaced the standard CNN encoder of the Retina U-Net architecture \cite{retina-unet} with a Swin Transformer \cite{2103.14030}, following a design similar to Swin UNeTR \cite{2201.01266}. 

The code associated with this study can be accessed at the following link: \url{https://github.com/SimoneBendazzoli93/nnDetection}

\section{Materials and Methods}

\subsection{Datasets}

In this study, we included two PET/CT datasets. The first, an open-access public dataset containing PET/CT scans with various annotated cancer lesions, was used for the self-supervised pretraining of the Swin Transformer. The second, a private collection acquired and annotated at Karolinska University Hospital, was used to train the detection model for the lymphoma lesion detection task.

\subsubsection{AutoPET Dataset}

\begin{figure}
    \centering
    \includegraphics[width=0.95\linewidth]{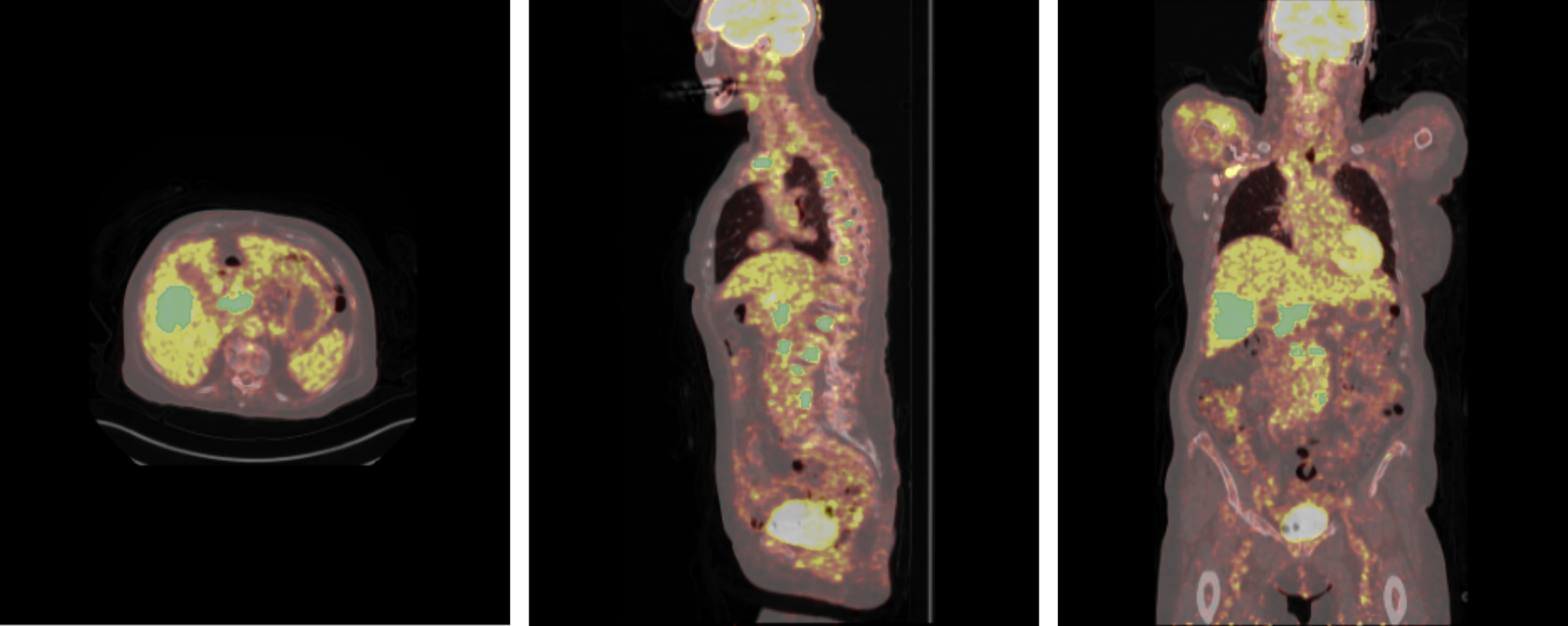}
    \caption{PET/CT scan from the AutoPET Dataset, with manual annotations of Lymphoma lesions highlighted in green. From left to right: Axial, Coronal and Sagittal views.}
    \label{fig:pet-ct}
\end{figure}

The first data set used in this study is the AutoPET dataset\cite{autopet,autopetdata}, This data set was used for the initial self-supervised pretraining stage of the Swin Transformer in our framework. The dataset comprises 1,611 whole body PET/CT scans (an example is shown in Figure~\ref{fig:pet-ct}). These include data from patients with histologically confirmed malignant melanoma (n=192), lymphoma (n=155), prostate carcinoma(n=332) or lung cancer (n=193), as well as a set of negative controls (n=739).  PET acquisitions were performed using three different tracers: 597 prostate-specific membrane antigen ([$^{18}$F]PSMA (n=369) or [$^{68}$Ga]PSMA(n=228)) scans and 1,014 fluorodeoxyglucose ([$^{18}$F]FDG).

Given the focus of this study on lymphoma lesion detection, we selected only the 872 PET/CT volumes with annotated cancer lesions for further use. Specifically, images from patients with lung cancer and melanoma, as well as prostate cancer were included, given the specific purpose of this dataset to provide self-supervised information on how to extract lesion features from PET/CT. 
Each case includes a 3D whole-body PET volume, a corresponding 3D whole-body CT scan, and a 3D binary mask of manually segmented tumor lesions. The FDG PET/CT data were annotated by a radiologist experienced in hybrid imaging, while the PSMA PET/CT data were annotated by a single reader and subsequently reviewed by a radiologist with expertise in hybrid imaging. PET and CT data were acquired sequentially on state-of-the-art hybrid PET/CT scanners (Siemens Biograph mCT for FDG-PET/CT, Siemens Biograph 64-4R TruePoint, Siemens Biograph mCT Flow 20, and GE Discovery 690 for PSMA-PET/CT) at two sites. The use of hybrid imaging ensures anatomical alignment, with only minor discrepancies potentially arising from physiological motion.

The scan range typically extended from the skull base to the mid-thigh, although in clinically indicated cases whole-body coverage was performed. For FDG-PET, an average dose of approximately 350~MBq of \textsuperscript{18}F-FDG was administered intravenously, with imaging initiated about 60 minutes post-injection. For PSMA-PET, imaging commenced on average 74 minutes after intravenous administration of 246~MBq \textsuperscript{18}F-PSMA (mean, 369 studies) or 214~MBq \textsuperscript{68}Ga-PSMA.  

Diagnostic contrast-enhanced CT scans were subsequently acquired, covering the neck, thorax, abdomen, and pelvis.
The CT slice thickness corresponding to FDG-PET was 2--3~mm. For PSMA-PET, CT slice thickness was 3~mm for the Siemens Biograph 64 and Biograph mCT systems, and 2.5~mm for the GE Discovery 690.

\subsubsection{Indolent Lymphoma KUH Dataset}
We also incorporated a second PET/CT dataset collected at Karolinska University Hospital (KUH), comprising patients diagnosed with various subtypes and stages of lymphoma.
This dataset was used in the second stage of our pipeline, focusing on training and evaluating the lymphoma lesion detection model.

A total of 228 whole-body PET/CT studies were included, with lymphoma lesions manually annotated by expert radiologists. Each study was manually reviewed and the lesion delineated: A radiologist annotated 151 cases, while the remaining 77 were annotated by a second radiologist, using different annotation tools. All patients had biopsy-confirmed lymphoma, and lesion boundaries were defined based on expert interpretation of the PET/CT images. PET and CT volumes were acquired on hybrid state-of-the-art PET/CT scanners (Siemens Biograph 128, Siemens Somatom Definition AS, GE Medical Systems Discovery) from two different sites. 

The scan range typically extended from the skull base to the mid-thigh, although whole-body coverage was performed in clinically relevant cases. Patients received an intravenous injection of on average 230 MBq of $^{18}$F-FDG, and PET acquisition was initiated approximately 70 minutes after injection.

For this study, attenuation correction CT (ACCT) was considered the primary modality of CT under investigation, since only 23 studies included diagnostic CT. For the Siemens Biograph and Siemens SOMATOM scanners, the CT slice thickness was 5 mm, while for the GE Discovery system it was 3.75 mm.

\subsection{Deep Learning Framework and Network Architectures}

\subsubsection{nnDetection Framework}
For the object detection task, we built on \texttt{nnDetection}~\cite{baumgartner2021nndetection}. \texttt{nnDetection} is the state-of-the-art for object detection in medical imaging. Built on the same principles as \textit{nnU-Net}~\cite{Isensee2020}, \texttt{nnDetection} is a self-configuring framework specifically designed for the detection of medical objects. The framework is designed to automatically select a range of hyperparameters, training configurations, and validation strategies through rule-based heuristics, based on the characteristics of the data set under investigation and the objects of interest, in our case, lymphoma lesions. \texttt{nnDetection} generates that way a fingerprint of the data set that is then used to optimize the architecture and hyperparameters so that the neural network is optimal for the task at hand.

In our case, the fingerprint of the dataset under analysis with \texttt{nnDetection} includes the spacing, shape, modality, intensity distribution, and number and size of the objects. Based on this fingerprint, several parameters are tuned to optimize the training and inference processes. These include the choice between full-resolution or cascade training (i.e., low-resolution followed by high-resolution stages), the batch size and patch size, as well as the resampling and normalization strategies used during data preprocessing.

\begin{figure}
    \centering
    \includegraphics[width=1\linewidth]{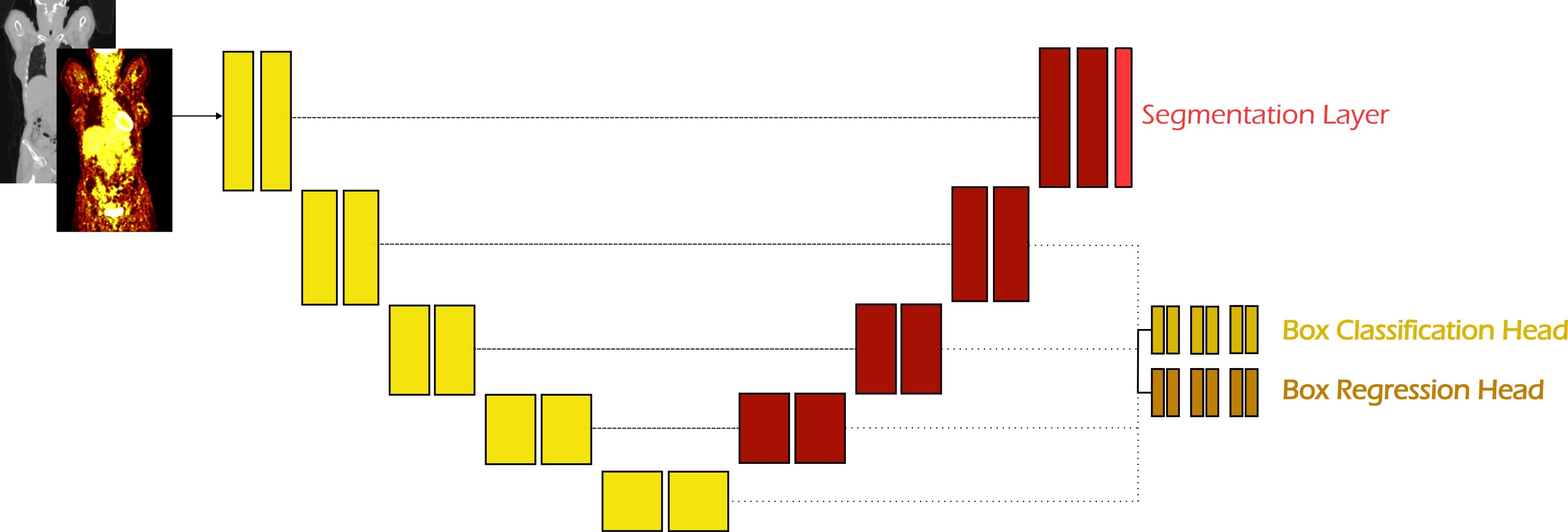}
    \caption{Retina U-Net architecture. The network consists of a 5-stage convolutional encoder (in yellow), followed by a convolutional  Feature Pyramid Network, serving as a decoder (in red). An auxiliary segmentation layer is attached to the highest-resolution decoder level to support the auxiliary segmentation task. Two parallel detection heads, one for box classification and one for box regression, are connected to the four lower decoder levels to enable multi-scale object detection, combining the different spatial resolutions of each decoder level.}
    \label{fig:retinaunet}
\end{figure}

The \texttt{nnDetection} framework adopts the Retina U-Net architecture. As illustrated in Figure \ref{fig:retinaunet}, Retina U-Net is a one-stage object detection model, structurally similar to RetinaNet \cite{1708.02002}, and specifically designed to address challenges such as class imbalance and multiscale object detection in medical imaging. \texttt{nnDetection} uses the fingerprint of the dataset to optimize the architecture. 
Additional parameters are tuned during the empirical optimization performed in the validation phase, including the non-maximum suppression (NMS) threshold, the minimum object size (to filter out small detections), the intersection over union (IoU) threshold, and the minimum model confidence score required to classify prediction boxes.

Retina U-Net utilizes anchor boxes, predefined bounding boxes with varying sizes and aspect ratios, to detect objects at different locations and scales. A significant innovation introduced in RetinaNet and also incorporated in Retina U-Net is the \textbf{ focal loss}. This loss function is specifically designed to tackle class imbalance by placing greater emphasis on hard-to-classify examples, achieved by focusing training on hard negative boxes, which are those that the model fails to detect.
The backbone network, typically a convolutional neural network, in this case a U-Net encoder, serves as the foundation for extracting feature maps from the input images. Subsequently, these feature maps are processed by a \textbf{ feature pyramid network (FPN)}, which generates a multiscale feature pyramid. The FPN integrates a bottom-up pathway, which involves upsampling coarser feature maps, alongside lateral connections that merge these upsampled maps with matching finer feature maps from the encoder.

Two subnets operate on the FPN's output at each stage, excluding the highest resolution one: a \textit{ classification head} and a \textit{regression head}. The classification head predicts the probability that an object is present at each spatial location for predefined anchor boxes and object classes, while the regression head refines the bounding-box coordinates by predicting a center offset and shape ratio for each predefined anchor box. Both subnets use convolutional layers to process the feature maps produced by the FPN.
A key difference between Retina U-Net and RetinaNet is that the highest resolution stage of the FPN in Retina U-Net is also used for an auxiliary segmentation task. According to the original paper, the gradient flow from this auxiliary task during backpropagation benefits the lower resolution stages involved in the object detection task.

\subsubsection{Inclusion of Anatomical Information}

\begin{figure}
    \centering
    \includegraphics[width=1\linewidth]{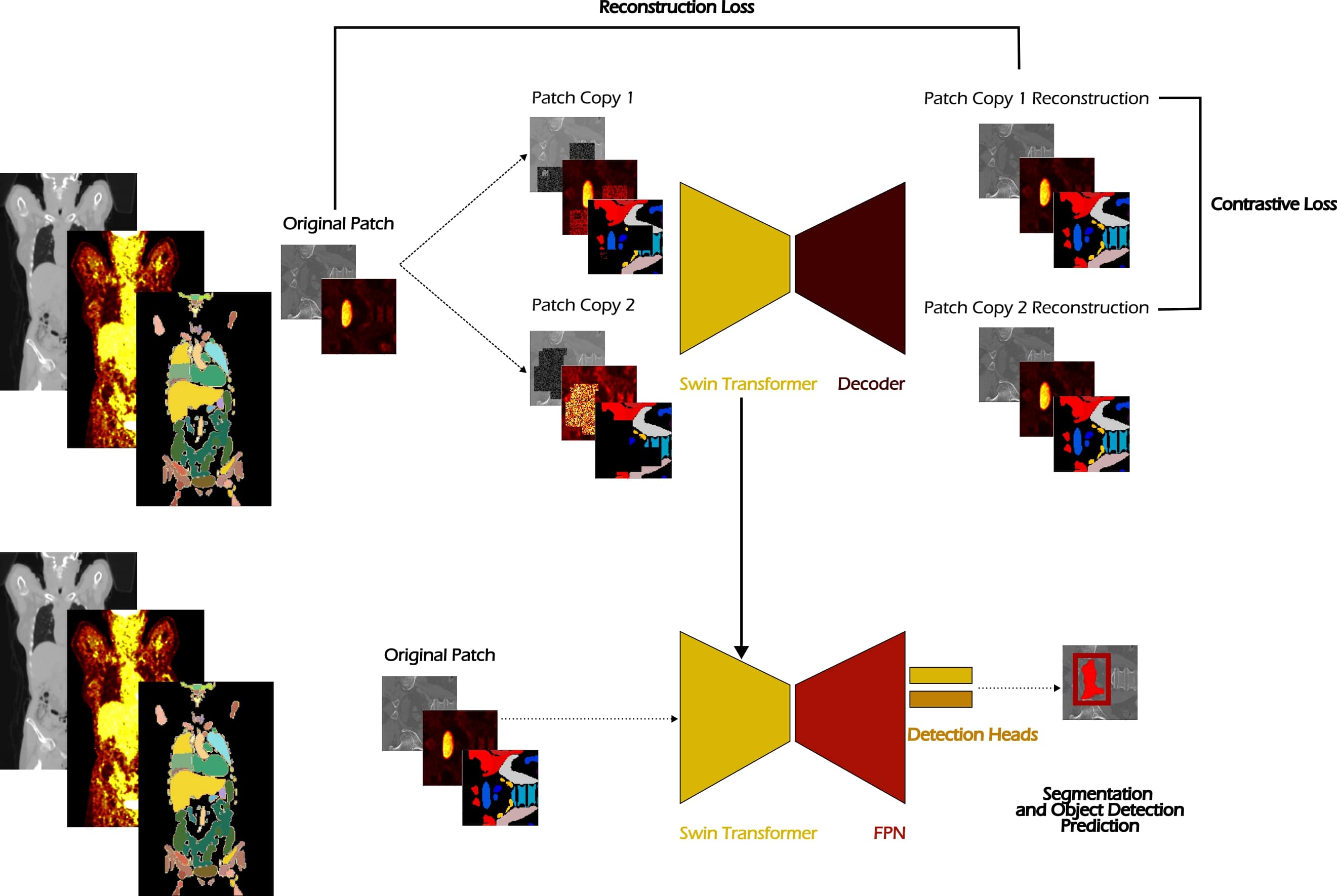}
    \caption{Overview of the experimental workflow integrating anatomical information via TotalSegmentator predictions. The generated segmentation masks are utilized in both the self-supervised pretraining phase, where they are included as input alongside corrupted PET/CT patches, and in the subsequent object detection training phase, where they are provided as additional input channels to the network.}
    \label{fig:plan-ts}
\end{figure}

To assess the impact of incorporating anatomical information as prior knowledge, the experimental procedure described above was performed in two variants. In the first variant, only PET and CT volumes were provided as input to the network, both during the self-supervised pre-training and the subsequent object detection training phase (see Figure~\ref{fig:plan-ts}). In the second variant, a third input channel was added, consisting of anatomical segmentation masks generated by the \textit{TotalSegmentator} tool. These masks, which cover 104 anatomical structures, were obtained by performing inference on the CT modality alone and served as an additional source of anatomical context.

\subsubsection{Self-Supervised Pretraining of Swin Transformer} 

ViTs, including the Swin Transformer, are typically large models with a substantial number of parameters, which makes them computationally intensive to train. A common strategy to address this challenge is to apply self-supervised pre-training on the transformer component, effectively "warming up" the model before full training. This practice was introduced and effectively demonstrated, notably in \cite{2111.14791}, where Swin UNeTR was pre-trained using proxy tasks before fine‑tuning on segmentation benchmarks. A related study by Xie et al. \cite{2105.04553}, presents the MoBY framework that adapts MoCo-v2\cite{2408.03230} and BYOL\cite{2006.07733} for Swin transformers and shows strong transfer performance in ImageNet and downstream segmentation and detection tasks.

Thus, for this initial self-supervised stage, we used the MONAI framework \cite{2211.02701}, which offers a rich collection of tutorials and modular components designed for deep learning in medical imaging. It includes numerous community-driven implementations of state-of-the-art architectures, image processing, and augmentation transforms, as well as utilities for training and validation.

\begin{figure}
    \centering
    \includegraphics[width=1\linewidth]{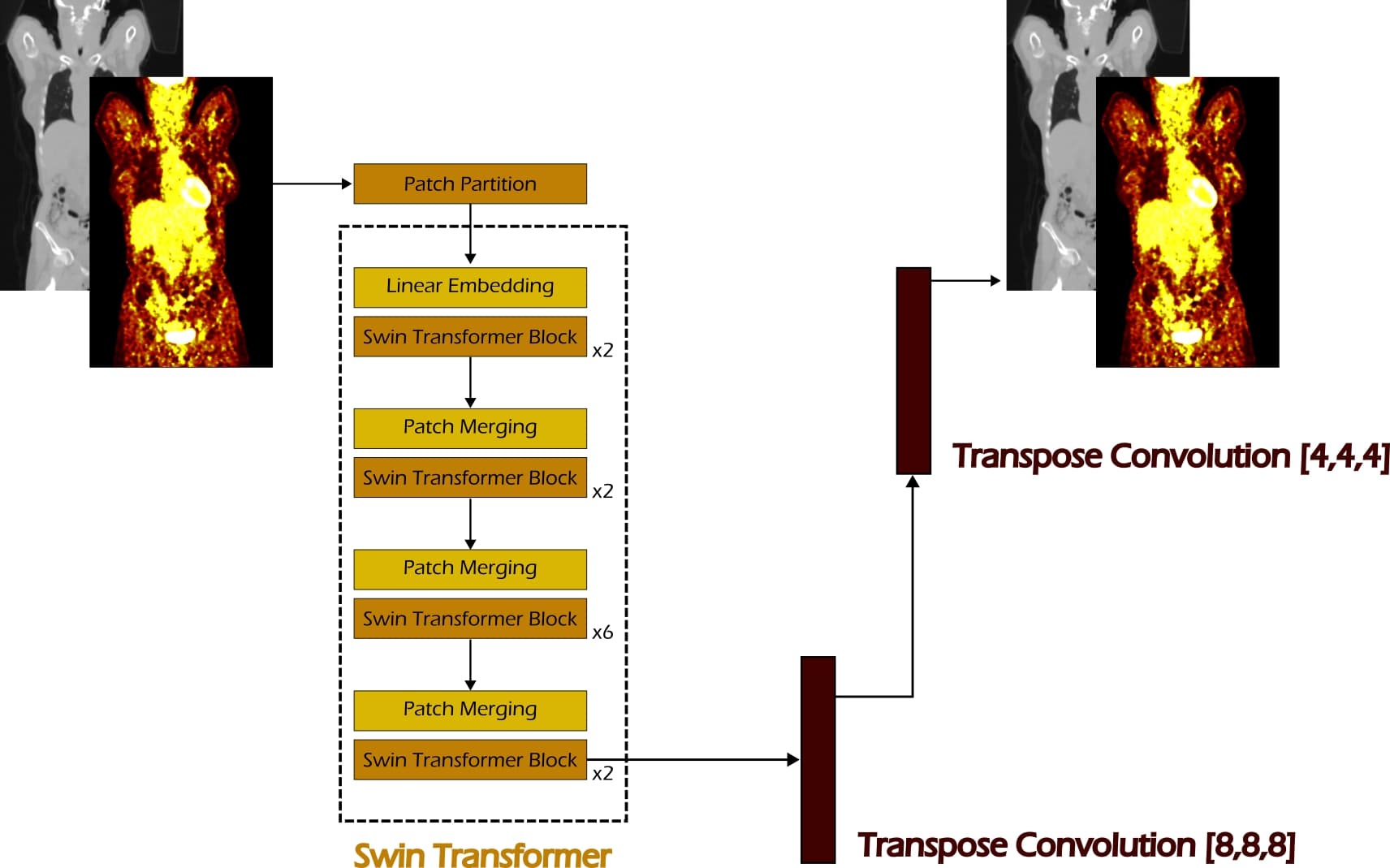}
    \caption{The proposed Swin Autoencoder network for the self-supervised task. A Swin Transformer is used as the feature extractor for the input PET/CT volumes, followed by a sequence of two transposed convolution layers to upsample the extracted features back to the original image resolution.}
    \label{fig:swin_autoencoder}
\end{figure}

As shown in Figure~\ref{fig:swin_autoencoder}, building on MONAI’s implementation of the Swin Transformer and drawing inspiration from self-supervised learning strategies used with ViT~\cite{2010.11929}, we extended the architecture by connecting the Swin Transformer bottleneck to a decoder composed of two transposed convolution layers, forming an autoencoder (see Figure~\ref{fig:swin_autoencoder}). This design enabled the model to be trained on a self-supervised image reconstruction task.
The Swin Transformer used in our setup consists of four stages, with the number of Swin Transformer blocks set to 2, 2, 6, and 2, respectively. The embedding dimension $C$ is set to 96, corresponding to the Swin-T architecture as defined in the original Swin Transformer paper \cite{2103.14030}. The window size is set to $4 \times 4 \times 4$, and the number of attention heads for each stage is 3, 6, 12, and 24, respectively.
 
The learning objective was to reconstruct corrupted input volumes, inspired by the masked autoencoding strategy described in~\cite{2111.14791}. A randomly cropped 3D patch from the input volume is duplicated, and each of the two copies is independently corrupted using a combination of the following transformations:

\begin{itemize}
    \item \textbf{Coarse Dropout}: Random rectangular regions were replaced with a random value in the range $[0, 0.2]$ or retained while the remaining areas of the image were filled with random values in the range $[0, 0.2]$, as described in~\cite{1708.04552,1604.07379}.
    \item \textbf{Pixel Shuffling}: Random regions within the image were selected and the pixels within each region were shuffled independently per channel~\cite{1707.07103}.
\end{itemize}

The reconstruction task was optimized using a combined loss function composed of a reconstruction loss term ($\mathcal{L}_{\text{rec}}$) and a contrastive loss term ($\mathcal{L}_{\text{con}}$) based on~\cite{pmlr-v119-chen20j}, with a temperature parameter of 0.05. The total loss was defined as:

\begin{equation}
    \mathcal{L}_{\text{total}} = \mathcal{L}_{\text{rec}} + \mathcal{L}_{\text{con}} \cdot \mathcal{L}_{\text{rec}}
\end{equation}

Training was performed for 500 epochs using the Adam optimizer with a fixed learning rate of $1 \times 10^{-4}$. The input volumes were randomly cropped into patches of size $96 \times 96 \times 96$, and a batch size of 2 was used.

From the available 872 PET/CT volumes with annotated cancer lesions in the AutoPET dataset, 698 were used for training and 174 for validation in a single train-validation split.

\subsubsection{Swin RetinaUNeTR}\label{swinretinaunetr}
\begin{figure}
    \centering
    \includegraphics[width=1\linewidth]{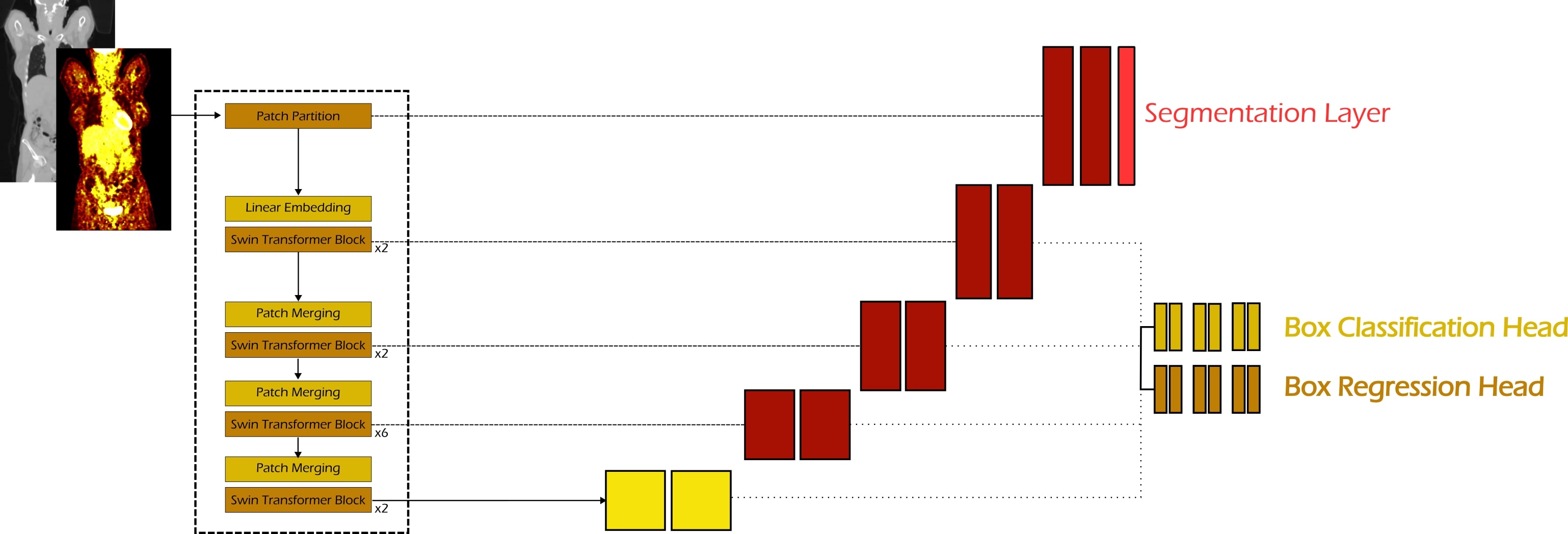}
    \caption{Proposed Swin RetinaUNeTR architecture. While maintaining structural similarity to Retina U-Net in the decoder and detection head components, the convolutional encoder is replaced by a Swin Transformer, serving as a vision transformer-based feature extractor.}
    \label{fig:retina-swinunetr}
\end{figure}

Based on the two primary architectures, Retina U-Net and the Swin Transformer, we assess the performance of ViT-based encoders on feature extraction for lymphoma lesion detection. In this model, the U-Net encoder of Retina U-Net is replaced by a Swin Transformer, as illustrated in Figure \ref{fig:retina-swinunetr}. The outputs from the first three stages, along with the output following the patch embedding at the original image resolution, are fed into the feature pyramid network to extract multiscale features. This design also inherits the Swin Transformer's ability to capture long-range self-attention features.

Thus, following the self-supervised pretraining phase, the Swin Transformer encoder was integrated with a Feature Pyramid Network (FPN) and detection heads within the \texttt{nnDetection} framework to train the Swin RetinaUneTR architecture for the lymphoma lesion detection task (see Figure~\ref{fig:retina-swinunetr}).

Training hyperparameters were automatically selected by \texttt{nnDetection}, based on a fingerprint extracted from the input data set. The multitask objective involved a combination of loss functions: Dice and cross-entropy losses for the segmentation component, Generalized Intersection over Union (GIoU) loss for bounding box regression, and standard Cross-Entropy loss for bounding box classification:

\begin{equation}
\mathcal{L}_{\text{total}} = \underbrace{\mathcal{L}_{\text{Dice}}^{\text{seg}} + \mathcal{L}_{\text{CE}}^{\text{seg}}}_{\text{Segmentation}} + \underbrace{\mathcal{L}_{\text{GIoU}}^{\text{reg}}}_{\text{Box Regression}} + \underbrace{\mathcal{L}_{\text{CE}}^{\text{cls}}}_{\text{Box Classification}}
\end{equation}

Model training was performed for 50 epochs, followed by an additional epoch of Stochastic Weight Averaging (SWA) \cite{1803.05407}. Each epoch consisted of 2500 iterations, using a batch size of 2 and a patch size of $96 \times 96 \times 96$ voxels.

The data augmentation strategy included a rich set of geometric and intensity-based transformations. Geometric augmentations comprised random elastic deformations, 3D rotations along all spatial axes, isotropic or anisotropic scaling, and random mirror flips across specified axes. Intensity enhancements included the application of Gaussian noise (10\% probability), Gaussian blurring with varying sigma values (applied to 20\% of samples), brightness scaling (multiplicative range $[0.75, 1.25]$ with 15\% probability) and optional additive brightness perturbations sampled from a Gaussian distribution. Further intensity manipulations included contrast adjustment and gamma correction, encompassing both standard and inverted gamma transformations, with configurable per-sample probabilities.

Stochastic gradient descent (SGD) was used as optimizer, with an initial learning rate of 0.01, momentum of 0.9, Nesterov acceleration enabled and a weight decay factor of $3 \times 10^{-5}$. The scheduling of the learning rate followed a warm polynomial decay strategy: the warm phase lasted 4,000 iterations starting from an initial learning rate of $1 \times 10^{-6}$, followed by a polynomial decay with a gamma parameter of 0.9.

\section{Results}

In the experiments, we used \texttt{nnDetection} with Retina U-Net as the baseline. This method was trained using the same procedure described in Section~\ref{swinretinaunetr}, using the \texttt{nnDetection} framework.
For both Swin RetinaUNeTR and baseline \texttt{nnDetection} training in the object detection task, a total of 228 PET/CT volumes with annotated cancer lesions from the KUH dataset were available. Of these, 182 were used for training and 46 for validation in a single train-validation split.

\begin{figure}
    \centering
    \includegraphics[width=0.75\linewidth]{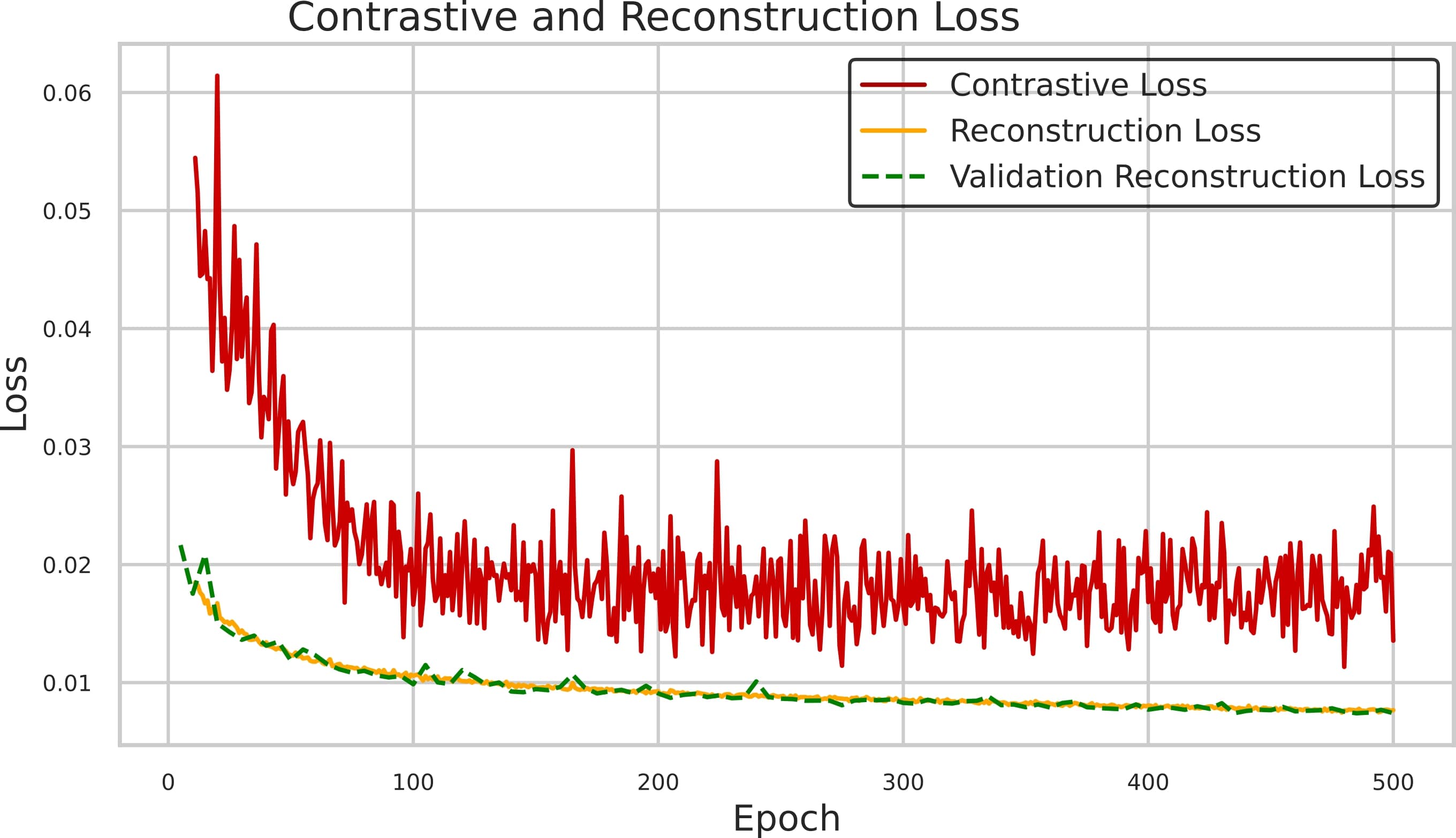}
    \caption{Training and validation loss curves across epochs. The training loss is split into contrastive and reconstruction components, shown separately. For validation, only the reconstruction loss is computed and displayed.}
    \label{fig:selfsup_curves}
\end{figure}

\begin{figure}
    \centering
    \includegraphics[width=0.8\linewidth]{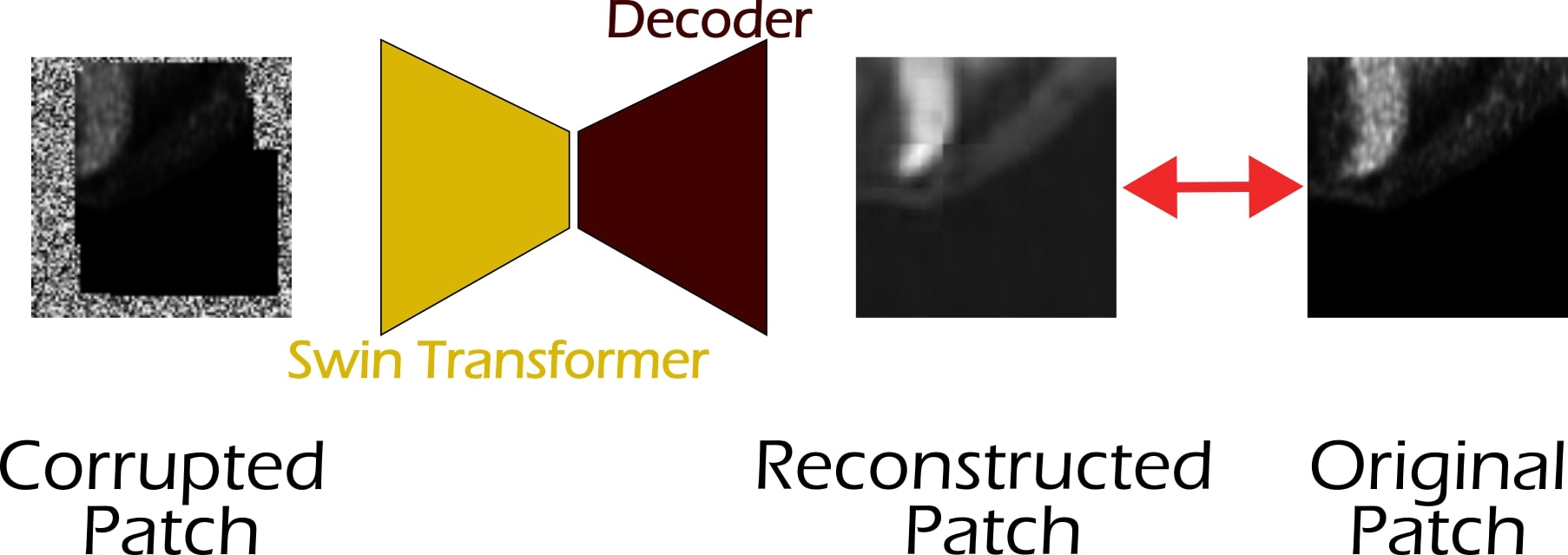}
    \caption{A randomly cropped $96 \times 96 \times 96$ sample from the validation set, corrupted during augmentation and subsequently reconstructed by the Swin autoencoder network.}
    \label{fig:selfsup_example}
\end{figure}

As illustrated in Figure~\ref{fig:selfsup_curves}, the training and validation curves for the self-supervised learning task indicate a stable and satisfactory convergence. The autoencoder effectively learned the reconstruction objective, as evidenced by the consistent decrease in both the reconstruction loss (\textit{L1 loss}) and the contrastive loss during training. For the validation set, only the reconstruction loss is shown.
Furthermore, Figure~\ref{fig:selfsup_example} presents a representative example of the validation set. The input patch, sized 96$\times$96$\times$96, was corrupted using random pixel shuffling and coarse dropout. The autoencoder reconstruction is compared with the original uncorrupted patch, demonstrating the model’s ability to recover a meaningful structure from perturbed input data.

For the object detection task, we use Average Precision (\textbf{AP}) and the Free-response Receiver Operating Characteristic (\textbf{FROC}) score as evaluation metrics, following the recommendations of the \textit{Metrics Reloaded} \cite{MaierHein2024} framework. Both metrics are well-suited for object detection and instance segmentation tasks, and belong to the family of multithreshold metrics, capturing performance across varying decision thresholds.

\begin{table}[ht]
\centering
\caption{Object detection performance across all experimental setups. \textbf{TS} refers to the inclusion of TotalSegmentator masks as additional input channels. Metrics reported are \textbf{FROC} at IoU thresholds 0.1 and 0.5, \textbf{AP} at IoU 0.1 and 0.5, and mean AP (\textbf{mAP}) averaged over IoUs from 0.1 to 0.5.}
\begin{tabular}{lccccc}
\hline
\textbf{Experiment} & \textbf{FROC@0.1} & \textbf{FROC@0.5} & \textbf{AP@0.1} & \textbf{AP@0.5} & \textbf{mAP@0.1--0.5} \\
\hline
nnDetection [Baseline]               & 0.284 & 0.115 & 0.418 & 0.110 & 0.288 \\
nnDetection + TS                     & \textbf{0.343} & \textbf{0.130} & \textbf{0.499} & \textbf{0.129} & \textbf{0.335} \\
Swin RetinaUNeTR                     & 0.243 & 0.074 & 0.372 & 0.065 & 0.237 \\
Swin RetinaUNeTR + TS                & 0.244 & 0.077 & 0.376 & 0.070 & 0.234 \\
\hline
\end{tabular}
\label{tab:detection_comparison}
\end{table}

\begin{figure}
    \centering
    \includegraphics[width=0.8\linewidth]{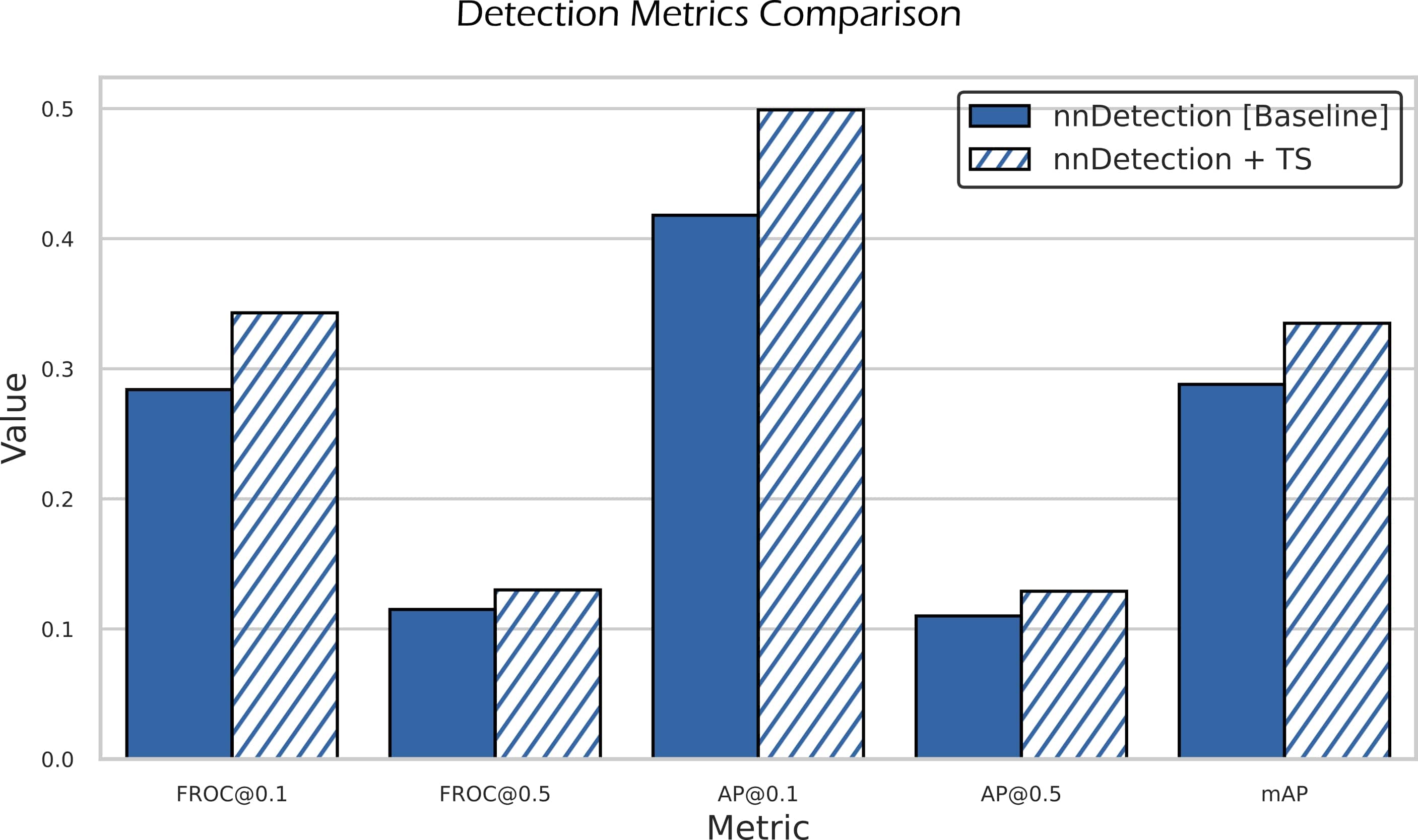}
    \caption{Comparison of detection metrics between the baseline nnDetection model and its enhanced version that incorporates TotalSegmentator masks as additional input. Shown metrics include FROC@0.1, FROC@0.5, AP@0.1, AP@0.5, and mAP.}
    \label{fig:nnDetection-TS}
\end{figure}

Table~\ref{tab:detection_comparison} summarizes the performance of four experimental setups evaluated using key object detection metrics: \textbf{FROC@0.1}, \textbf{AP@0.1}, and \textbf{mAP}. Among them, the \texttt{nnDetection} + TS model, where TotalSegmentator anatomical masks are added as input channels, achieves the best results in all three metrics, with a FROC@0.1 of 0.343, AP@0.1 of 0.499, and mAP of 0.335. These values represent relative gains of 0.059, 0.081, and 0.047, respectively, over the \texttt{nnDetection}  baseline (FROC@0.1 = 0.284, AP@0.1 = 0.418, mAP = 0.288), as shown in Figure~\ref{fig:nnDetection-TS}.

\begin{figure}
    \centering
    \includegraphics[width=0.8\linewidth]{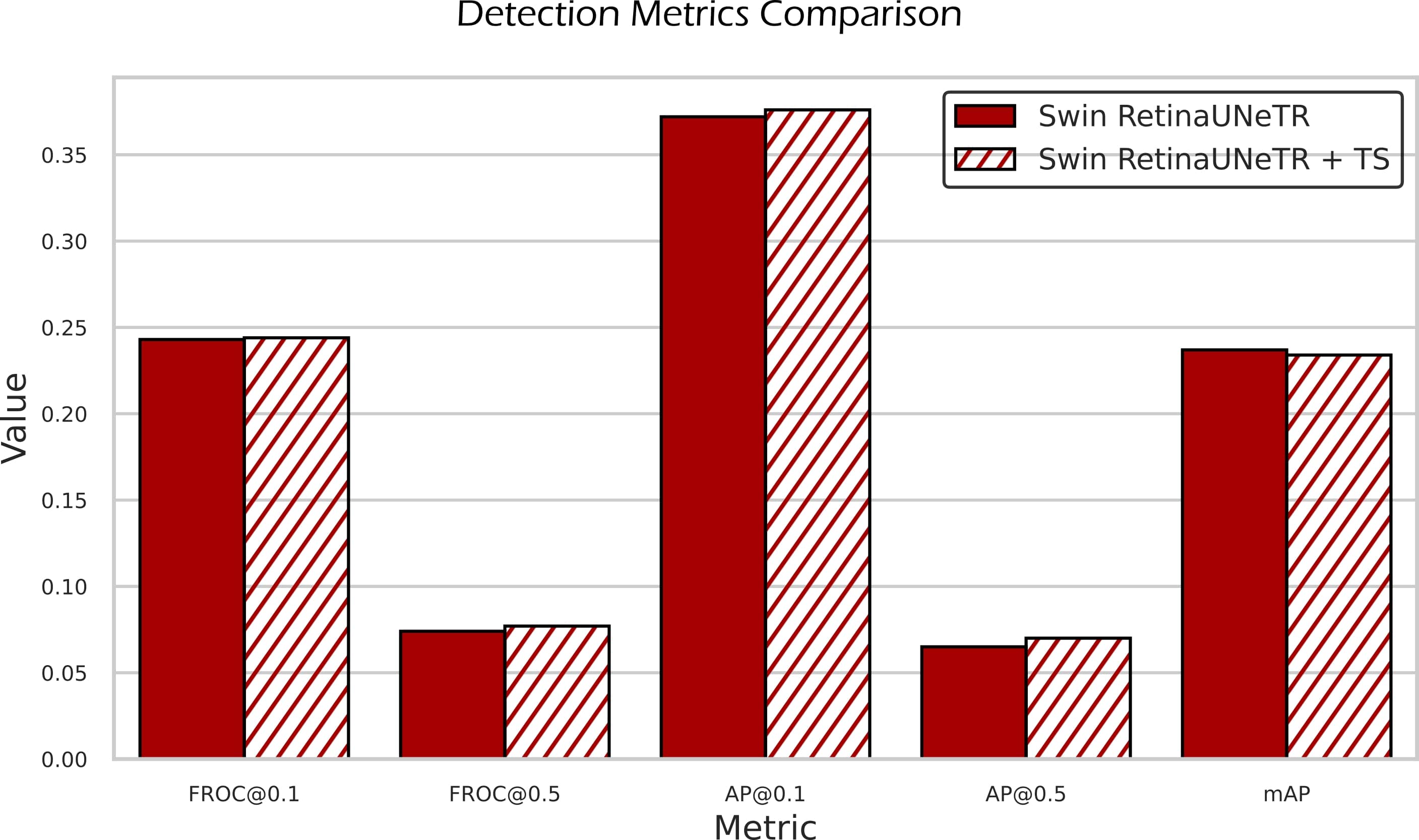}
    \caption{Comparison of detection metrics between the proposed Swin RetinaUNeTR model and its enhanced version that incorporates TotalSegmentator masks as additional input. Shown metrics include FROC@0.1, FROC@0.5, AP@0.1, AP@0.5, and mAP.}
    \label{fig:Swin-TS}
\end{figure}

In contrast, as illustrated in Figure~\ref{fig:Swin-TS}, adding the same anatomical information to the Swin RetinaUNeTR model produces only marginal gains, resulting in 0.244 (FROC@0.1), 0.376 (AP@0.1) and 0.234 (mAP), which are nearly identical to the baseline performance of Swin RetinaUNeTR.

\begin{figure}
    \centering
    \includegraphics[width=0.8\linewidth]{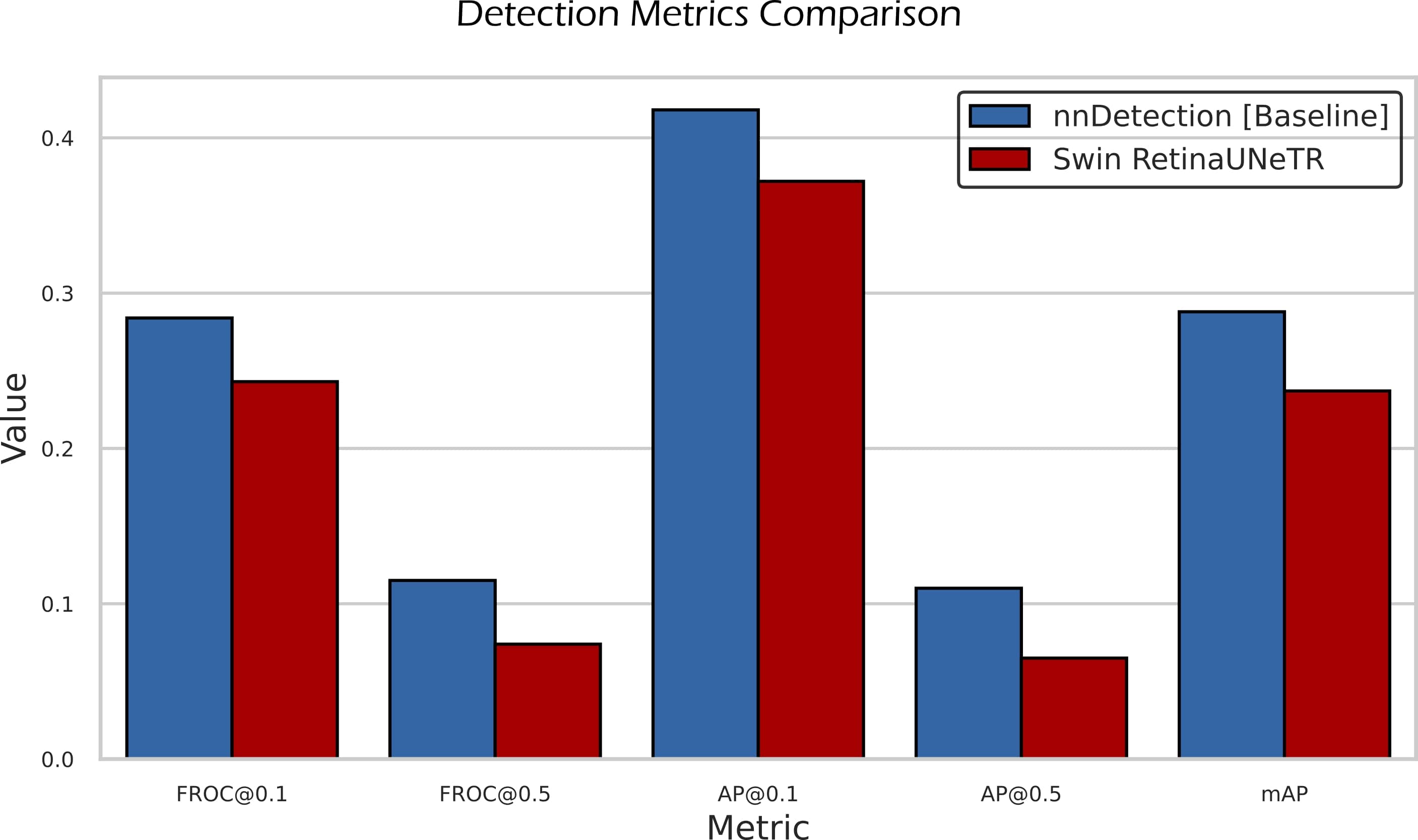}
    \caption{Comparison of detection metrics between the baseline nnDetection model and the proposed Swin RetinaUNeTR model. Shown metrics include FROC@0.1, FROC@0.5, AP@0.1, AP@0.5, and mAP.}
    \label{fig:nnDetection-Swin}
\end{figure}

When directly comparing the Swin RetinaUNeTR model to the \texttt{nnDetection}  baseline (see Figure~\ref{fig:nnDetection-Swin}), the latter clearly demonstrates superior performance in all three detection metrics. The \texttt{nnDetection}  baseline surpasses Swin RetinaUNeTR with a FROC@0.1 of 0.284 vs. 0.243, an AP@0.1 of 0.418 vs. 0.372, and a mAP of 0.288 vs. 0.237. These results indicate that \texttt{nnDetection}  offers more robust low-threshold sensitivity and precision, as well as better overall localization performance across the IoU range.

In summary, integrating anatomical priors through TotalSegmentator masks consistently enhances the performance of the \texttt{nnDetection}  framework, making it the most effective strategy among all tested configurations. In contrast, the Swin RetinaUNeTR architecture shows minimal responsiveness to the same anatomical input, suggesting that its ViT-based design may not effectively integrate and take advantage of the semantic anatomical information provided for the lesion detection task in this context.

\section{Discussion}
In this study, we investigated the effect of adding anatomical priors for identifying lymphoma lesions in whole-body PET/CT scans. For this, we incorporated anatomical information into the training pipeline using TotalSegmentator to generate segmentation masks of 104 organs from the CT modality. The masks were then provided as additional input channels in conjunction with the PET and CT volumes. Using the  \texttt{nnDetection} framework as a baseline, we extended the Retina U-Net architecture by explicitly incorporating anatomical information as an additional input during training. Moreover, we adapted the Swin transformer architecture for lesion detection and assessed the effect of including extra anatomical information in that model. 

\begin{figure}
    \centering
    \includegraphics[width=0.75\linewidth]{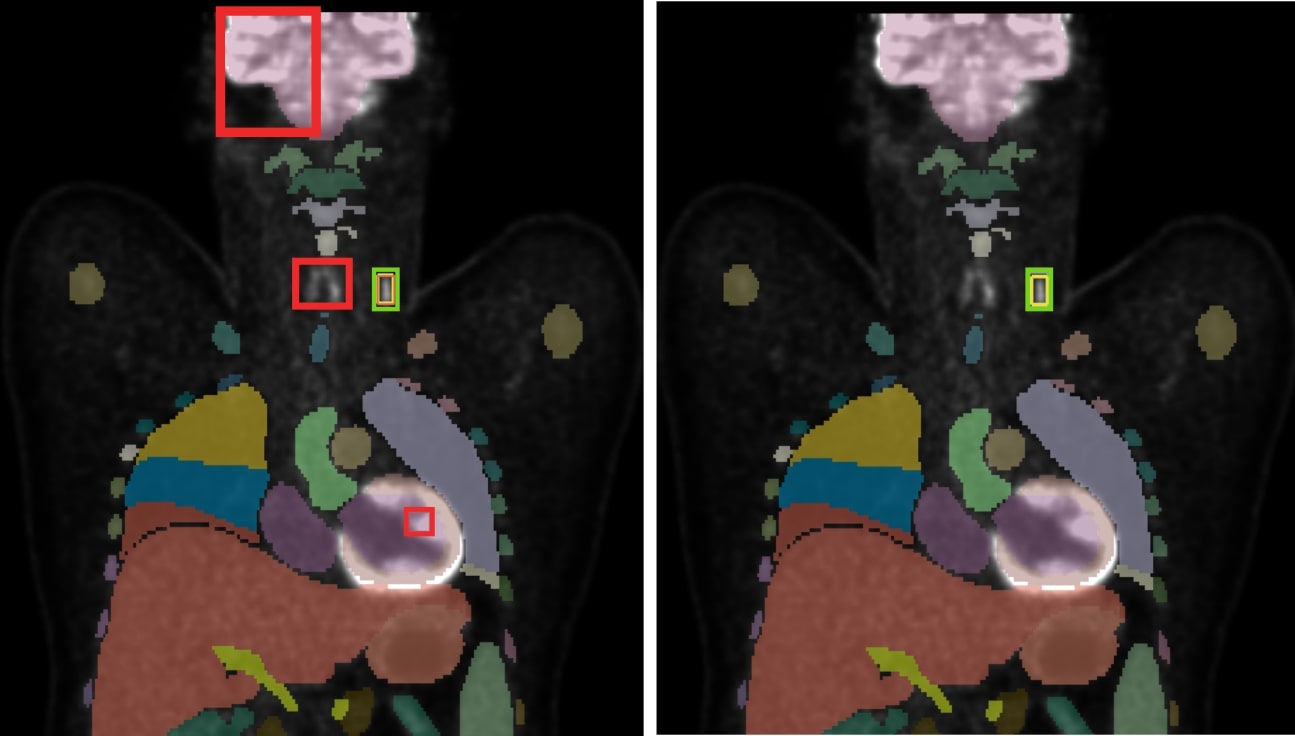}
    \caption{Visualization of detection results: baseline nnDetection (left) vs. enhanced nnDetection (right) with Total Segmentator mask overlay. The incorporation of anatomical information in the enhanced model demonstrates a reduction in false positive detections, particularly evident in the brain, thyroid, and heart regions.}
    \label{fig:ts-example}
\end{figure}

Experimental results demonstrated that the inclusion of anatomical information led to improved performance, particularly in the \texttt{nnDetection}  framework. These findings suggest that the network is capable of learning contextual anatomical information, correlating the occurrence of lymphoma lesions with specific organs, while also learning to ignore physiologically high-uptake regions in PET, such as the liver, thyroid, brain, and bladder, as shown in Figure~\ref{fig:ts-example}. This performance boost is likely attributable to the regionally localized nature of convolutional filters used in \texttt{nnDetection}, which appear to benefit from the explicit anatomical context provided.

In contrast, the same benefits were not observed in the Swin transformer-based architecture. The addition of anatomical input did not lead to significant performance improvements, possibly due to the global attention mechanism inherent in the Swin Transformer. This characteristic may already allow the model to extract contextual dependencies, reducing the marginal gain from the added anatomical information.

When directly comparing the Swin transformer-based network with the well-established \texttt{nnDetection}  framework, it is evident that the former does not yet match the state-of-the-art performance of the latter. This observation highlights two important conclusions: (1) \texttt{nnDetection}  continues to demonstrate robust, high-performing capabilities across diverse tasks and remains a strong benchmark in medical image object detection; (2) the Swin RetinaUNeTR architecture may require further refinement, particularly in the tuning and optimization of the Swin Transformer component, to realize its full potential and achieve competitive performance.

Our future work will focus on this refinement process and extend the evaluation of the proposed approach to additional medical object detection tasks beyond lymphoma, such as multifocal tumors in the liver or brain, and across different imaging modalities. Moreover, proposing better approaches for warming up Swin transformers may help to close the current gap between CNN-based and ViT-based lymphoma detection architectures. Since \texttt{nnDetection} is optimized for CNN architectures, it is relevant to propose alternative autoconfiguration mechanisms that can be more effective with ViT architectures at a reasonable cost.
Once transformer-based architectures achieve state-of-the-art performance and comparable robustness, the focus can shift toward multi-objective studies. For example, a single network could be leveraged in a multitasking framework to both detect lymphoma lesions and simultaneously classify disease subtypes, such as indolent versus aggressive lymphoma.
Finally, as presented in this study, attenuation correction CT (ACCT), which is acquired in standard procedures during PET for correction of photon attenuation of fat and other tissues, can be considered a feasible reference CT modality for future studies. This approach can eliminate the need for an additional contrast-enhanced diagnostic CT acquisition. However, further validation through comparative analyses with and without contrast-enhanced diagnostic CT remains necessary.

It is important to note that one of the primary objectives in the development of deep learning methods is to demonstrate tangible clinical benefits. Currently, we have not studied the influence of adding anatomical priors on reporting time, staging accuracy, or treatment decisions. It is also important to stress that this work focused exclusively on a single disease and, therefore, the results cannot be assumed to generalize to other cancer types without further investigation.

\section{Conclusion}
In conclusion, our results suggest that integrating anatomical information through TotalSegmentator predictions is effective for improving the performance of CNN-based lesion detection architectures. Adding such information has little effect on vision transformer architectures. 
Further developments are necessary to increase the performance of transformer-based architectures for lesion detection, particularly in light of the current limitations observed with Swin RetinaUNeTR.

\section*{Conflict of Interest Statement}

The authors declare that the research was conducted in the absence of any commercial or financial relationships that could be construed as a potential conflict of interest.



\section*{Funding}
This study has been partially funded by the Swedish Childhood Cancer Foundation (Barncancerfonden MT2022-0008), by Vinnova through AIDA, project ID: 2319,by the Swedish Research Council (Vetenskapsrådet, grant 2022-03389), and Hjärt-Lungfonden (grant No. 2022-0492). 

\section*{Acknowledgments}
We thank the National Academic Infrastructure for Supercomputing in Sweden (NAISS) for the computational resources at Alvis and PDC.


\section*{Data Availability Statement}
The use of Indolent lymphoma data was approved by the Regional Ethics Review Board in Stockholm under approval number 2012/783-31/3. 
An amendment clarifying the inclusion of X-ray imaging was approved in 2020 under the reference number 2019-06366.
An amendment was requested and approved for the use of lymphoma diagnoses under reference number 2016/2379-32.

\bibliographystyle{Frontiers-Vancouver} 

\bibliography{Bibliography}

\newpage
\section*{Figure captions}



\begin{figure}
    \centering
    \includegraphics[width=1\linewidth]{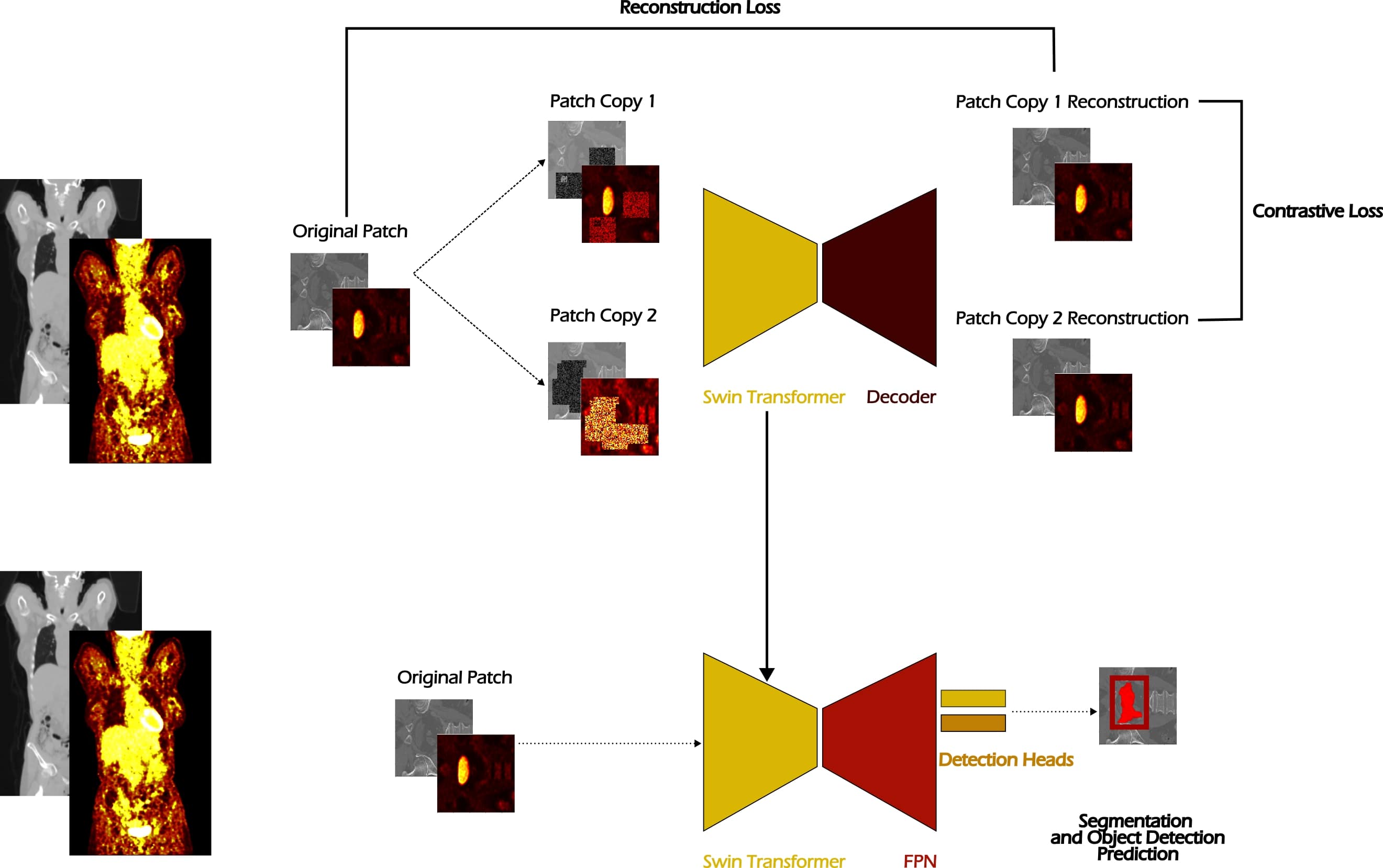}
    \caption{Overview of the experimental workflow. The process begins with a self-supervised pretraining phase, where random PET/CT patches are corrupted and reconstructed using a Swin Transformer-based autoencoder. This is followed by an object detection training stage, where the pretrained Swin Transformer is integrated into a Feature Pyramid Network (FPN) and detection heads for multi-scale object detection.}
    \label{fig:plan}
\end{figure}



\end{document}